# Heating of the Solar Wind by Ion Acoustic Waves

Paul J Kellogg
University of Minnesota

**Abstract**
Calculations are made of the energy supplied to the solar wind by the rapid decay of density fluctuations, identified as ion acoustic waves. It is shown that this process supplies an appreciable fraction, perhaps nearly all, of the observed heating of the solar wind. This process may be an important step in the conversion of magnetic turbulence to particle energy

## 1. Introduction

The plasma wind from the sun, the solar wind, is strongly accelerated near the sun, but this heating and acceleration continues at least to 1 AU. To understand this is, of course, a major goal of the recently launched Parker Solar Probe spacecraft, the second of missions to explore this heating process. It has been suggested (Tu and Marsch (1994,1995), Howes et al (2012), Versharen et al (2017), Narita and Marsch (2015)) that the ubiquitous density fluctuations in the solar wind are waves in the ion acoustic mode, aka kinetic slow mode. As these are quickly turned into particle energy, it is of interest to calculate how much heating is provided and the purpose of this note is to make this calculation.

According to Vlasov calculations, ion acoustic modes are very strongly damped, with an imaginary part of the frequency of the order of a third of the real part, (Barnes (1966)) thus giving a damping time of the order of half a cycle. If ion acoustic waves are an important part of solar wind turbulence this poses two problems, first why are there so many such waves when they are so evanescent and (second) is there too much heating of the solar wind? Several ways out of this perceived problem have been suggested. Howes et al (2012) suggested that the wave vector could be very oblique, so that both the real and imaginary part of the frequency are near zero, so that absorption is slow. Tu and Marsch (1994) suggested that the Vlasov calculations are correct and that the waves were in fact absorbed, and their energy contributed to the energization and acceleration of the solar wind. A third suggestion has been proposed by Schekochihin and colleagues, (2009) Howes et al (2006) and Parker et al (2016) that the Vlasov calculations do not give the correct answer for the damping in some cases.

In this note, I calculate and use in-situ observations to investigate whether there are enough zero frequency waves to account for very slow absorption. It seems there are not. However, the identification of density fluctuations with the ion acoustic mode implies that the energy delivered to the ambient plasma of the solar wind by their decay is, within uncertainties, consistent with the observations of heating at 1 A.U. It is therefor found that the perceived problem is no problem.

It seems that the suggestion of Schekochihin and colleagues (Schekochihin et al, (2009), Howes et al (2006),Parker et al (2016)) that the Vlasov damping is not correct in some cases is something that cannot be tested by comparison with observations, but perhaps by simulations.

## 2, Wave Modes

Partly for the author's own edification, a short discussion of wave mode designations follows. There are three popularly discussed wave modes in plasma. They have a dozen or two dozen names however. Further, in Vlasov theory, there are many more modes, but only some of them correspond to MHD modes. The most popular set of names comes from MHD. The dispersion relation, the vanishing of the determinant of the MHD equations, is:

$$\omega (\omega^2 - k_z^2 V_A^2)[\omega^4 - \omega^2 k^2 (V_A^2 + V_S^2) + k^2 k_z^2 V_A^2 V_S^2] = 0$$

The solution allows a fourth solution, corresponding to $\omega = 0$, which, like all $\omega = 0$ solutions, must be a pressure balanced mode. It is often ignored (though not by Verscharen et al 2017). It is called the entropy mode or the NP (non propagating) mode. The solution corresponding to the next expression in brackets will be called the shear Alfven mode here. The remaining expression, in square brackets, corresponds to two solutions. For propagation parallel to the steady state $B_0$, ($k_z = k$) these are: $\omega = kV_A$ and $\omega = kV_S$, where $V_A$ and $V_S$ are the Alfven speed and the ion sound speed. It is usual to call the solution corresponding to $\omega = kV_S$ and its extension to oblique propagation the slow mode, although if $\beta_w$, the square of the ratio $V_S/V_A$ is greater than 1, it is actually faster. The corresponding Vlasov solution is called the kinetic slow mode in recent literature. (The various authors have used different names for this mode). However, ion acoustic carries the suggestion that the energy of these waves is acoustic, which is true, but "kinetic" emphasizes that these waves are strongly damped whereas they are not in MHD, Roberts, (1966). In this work both names will be used. However, during much of the time used for the two analyses, the ion acoustic speed is faster than the slow speed, but the mode considered is still the highly damped mode in spite of the slow mode name. In early work and in the text books (e.g. Stix 1962, 1992), it is called the ion acoustic mode or the ion sound mode, but kinetic slow mode has become common in the recent literature and will also be used here. The solution corresponding to $\omega = kV_A$ for parallel propagation is called the electromagnetic ion cyclotron mode in magnetospheric work. It does not enter much in the discussion here.

## 2.1 $\omega = 0$ modes.

In addition to the $\omega = 0$ of the MHD equations, in the limit of extreme obliquity, i.e wave vector perpendicular to the ambient magnetic field, ion acoustic modes become pressure balanced modes, as must all $\omega = 0$ waves. Pressure balanced structure do occur in the solar wind (Burlaga and Ogilvie,(1970a,b), Tu and Marsch (1995), Kellogg and Horbury (2005), Yao et al (2011). In the solar wind, magnetic pressure is generally of the same order as particle pressure, but it appears that this is more equipartition than accurate pressure balance. In Figure 1 are shown two histograms of the observed ratio of magnetic pressure to particle pressure using data from the 3DP experiment (Lin et al,(1995)) and the MFI experiment (Lepping et al (1995)) on Wind. These histograms are from two periods that have been analyzed for this work, 2005 Feb 17, and 2005 Feb 5. It will be seen that accurately balanced pressures are sufficiently rare that it seems that they could not account for the common negative correlation between density fluctuations and the fluctuations of magnetic field parallel to the average magnetic field used by Howes et al (2012) to identify ion acoustic waves.

## 3. Energy in Waves and Their Absorption

### 3.1 Some Expressions for Wave Energy

According to the solutions of the Vlasov equations for propagation parallel to the magnetic field, the imaginary part of the frequency of ion acoustic waves is of the order of one third of the real part, leading to absorption times of half a cycle (Barnes 1966) Tu and Marsch suggested, assuming that this is correct, that the energy of the waves is converted to particle energy at this rate, leading to some heating and acceleration of the solar wind. More recently, Narita and Marsch (2015) have made an extensive analysis of the Vlasov dispersion relations for these ion acoustic waves, confirming this imaginary part of the frequency, but finding much change with angle of propagation. In this note, calculations of the expected rate of such heating are presented.

There are, in the literature, several expressions for the energy in such waves, Three different approaches to the energy and the energy transferred have been tried.. A common treatment of wave energy is proportional to the square of the electric field fluctuations (Auerbach 1979). In ion acoustic waves, the change in ion density almost exactly balances the change in electron density, a balance which becomes more and more perfect at lower frequencies, so that the electric field is small. Almost all of the energy is then in the acoustic system, not the electric field, which is why the name ion acoustic is mostly used here.

The common expression for the energy in a wave mode which relates the energy of electric field (Brillouin 1921, Landau and Lifshitz 1960, 1969, Auerbach 1979) is:.

$$W(\omega,k) = \omega \frac{\partial \varepsilon'}{\partial \omega} \frac{\langle E(x,t)^2 \rangle}{4\pi} \qquad (1)$$

Here $\varepsilon'$ is the real part of the derivative of Z, the well known electrostatic dispersion function (e.g. Fried and Conte 1961, Stix, 1962,1992) which for an electrostatic wave is the sum over species as below:

$$\varepsilon(k,\omega) = 1 - \sum_j \frac{1}{2k^2 \lambda_{Dj}^2} Z'(\zeta_j) \qquad (2)$$

Other relations are given by Landau and Lifshitz (1960,1969, Eq 61.4). They give an equation for the energy transferred, not the energy, from a wave with averaged electric field $E^2$ in the electrostatic limit. However, both of these approaches require knowledge of the mode, frequency and wavenumber of the wave being investigated. These are not known. A different approach is taken here.

### 3.2 Acoustic Energy

Therefor the energy in ion-acoustic-slow mode waves will be calculated using an expression for sound waves. The energy in a sound wave. i.e the energy in particles, (Rayleigh 1894) is

$$W = pressure + velocity = \int_V \frac{\delta p^2}{2\rho_0 V_S^2} dV + \int_V \frac{\rho_0 \delta v^2}{2} dV \qquad (3)$$

For sound the sum is, of course, constant but oscillates between terms. To evaluate the first term, the wave pressure energy at its maximum value, a set of data from the 3DP experiment (Lin et al, 1995) on the Wind mission is used. The purpose here is to evaluate the energy in ion acoustic waves, and the pressure is expected to follow the plasma density. The heating, the damping rate of the waves which is the imaginary part of the frequency, depends on frequency, so a set of measurements of some length must be used both to establish the average density and to establish the spectrum by Fourier transform. On the other hand, the presence of an average pressure in $\delta p$ of the expression, requires that the set not be too long, as then it might include effects such as discontinuities, compression regions, Langmuir waves from Type III bursts, etc. which would distort the results. This approach cannot lead to precise results. $\delta p$, the deviation from an average density, would be easy for the atmosphere, but the solar wind consists of many different plasmas, separated by current sheets and discontinuities. The choice of length is a compromise and limits the accuracy of the estimates.

As the calculations of Eq. (3) are to be compared with direct measurements, particularly those of Coleman(1968) a period in 2005 which is at the same phase of the solar cycle as was Mariner II and Coleman's measurement, has been chosen for the present analysis. A period of 14 hours, from 2005 Jan 17 1000 to 2400 was chosen, but then another period 2005 Jan 05 0900 to 2005 Jan 05 2400 was added. The first period was chosen at random, but it was found that the ion pressure was unusually large, and the second period was found by hunting for a period of low pressure. During most of these periods, the solar wind was slow, though some faster wind is found. During the Mariner II mission, there was an appreciable fraction of fast solar wind.

Figure 2 shows a spectrogram of the wave energy during the high density period. The data are not normalized to energy but are presented to show the exponent of the power. The average wave energy of a single observation over the whole period is $4.07 \; 10^{-13}$ J/m$^3$. These data may be fitted with power $\alpha \; f^{(-0.80)}$, shown as a red line. As it is known that the spectrum of density follows a -5/3 power law (Chen et al 2013, 2014 ) an exponent of -0.8 might cast some doubt on the identification of pressure as being due to ion acoustic waves  It seems there must be some correlation between density and temperature. This is shown in Figure 3, for the two, high and low pressure periods. The red curves are the expressions $1.6 \; \rho^{0.69}$ for electrons and $4.6 \; \rho^{0.56}$ for protons for the high pressure period, Feb 17. For the low pressure period, Feb 5, the curves are $3. \; \rho^{0.31}$ for electrons and $4.6 \; \rho^{0.53}$ for protons. It is expected that there be some difference between the exponents for electrons and ions, as the $\gamma$'s are often taken to be 3 for ions and 1 for electrons, but neither is found in this set. At any rate, the purpose here is to show that correlations alter the spectrum of wave energy from what might otherwise be expected to be similar to the density spectrum.

For further work, sets of 128 single observations, are chosen. The time for such a set, 128 times the observation cadence of 3.04 sec, is 389 sec, fairly close to the 300 sec. samples used by Howes et al (2012.)  The wave energies and the spectrum in each set is than evaluated from Eq. 3 and from an FFT of the 389 sec. of energy. The heating rate is then found from a convolution of the Fourier transform of the wave energy and the same transform of the damping rate, the imaginary part of the wave frequency. The inverse of the convolution in then a set of

128 measurements of the heating and their average is plotted in Figure 4 for each of the two periods analyzed.

In this development, $\omega$ is the angular frequency in the rest frame of the plasma, whereas the measurements to be used are often the observed frequency in the moving solar wind, which according to the Taylor hypothesis is $\omega_{obs} = k_{\|v} v_{sw}$. $k_{\|v}$ is the component of the wave vector in the direction of the solar wind. As $k_{\|v}$ is not measured, we use $k_{\|v} = 2 \pi f_{obs}/V_{sw}$ and also assume the dispersion relation for ion acoustic waves $\omega = k_{\|B} V_S$ where $V_S$ is the ion sound speed sqrt($(k_B T_e + 3 k_B T_p)/M_p$) and $k_{\|B}$ is the component of the wave vector in the direction of B.

The relation between $f_{obs}$ and $\omega$ is then $f_{obs} = (1/2\pi) (V_{sw}/V_s) \omega$. For both of the periods analyzed, the solar wind speed was within 20 km/s of $V_{sw}$ = 390 km/s, and this was used for the connection between f and $\omega$. Commonly in the data, $V_s \sim 80$ km/s, but the actual value was used. This typical value give the ratio, $f_{obs}/\omega = .8$, not much different from unity. On the average then, the damping is approximately f/4 /sec.

The results of this calculation of heating are shown in figure 4 for the low pressure and for the high pressure periods. The horizontal lines in these figures are the heating rates found by Coleman (1968) and by Gazis and Lazarus (1982) to be discussed below.

## .1.2 Comparison with Observations

Coleman (1968) found the heating of protons required 2.4 $10^6$ ergs/gm-s. For the observed average density of 5.6 ions/cm$^3$ or 9.4 $10^{-21}$ kg/m$^3$ this heating is 2.2 -18 J/m$^3$-s

More recently, heating has been obtained by fitting data from temperature as a function of distance from the sun:

$$T(R) = T_0 (R/R_0)^{-\alpha} \tag{4}$$

with various values of $\alpha$. Typical is $\alpha = .7$ ( e.g. Gazis and Lazarus 1982) For adiabatic cooling without other heating, $\alpha = 4/3$. For heating H J/sec per particle, the evolution of $k_B T$ would be: $k_B T/dR = (1/V_{SW}) k_B T/dt = H/ V_{SW}$, with the consequence that the heating per proton required to maintain a temperature exponent $\alpha$ is

$$H = k_B T ( 4/3 - \alpha) V_{SW}/R \tag{5}$$

For $\alpha = .7$ $V_{SW} = 450$ km/s and $k_B T = 5$ eV and R = 1 AU, this implies 9.9 $10^{-6}$ eV/s per sec,proton, or 8.9 10-18 J/m3 for the Mariner density. This is about 4 times higher that Coleman (1968)

The best mission for the study of heating within 1 AU is the Helios mission. The Helios data have been recently reworked ((Stansby et al 2018)) with different values of $\alpha$ for different components of the solar wind. The values are generally between .7 and 1., in accord with the results above.

The horizontal lines in Figure 4 show these measurements of heating, with the Coleman measurements being the lower. It will be seen that the calculated heating is frequently larger, sometimes one or two orders of magnitude larger than the observations. Three reasons have been found which account for this overestimate.

First, a part of this discrepancy is wave mixture. In this work, no certain identification has been made of ion acoustic waves. It is only thought that using the pressure part of Rayleigh's formula will be inclined toward ion acoustic waves. There is undoubtedly some mixture of other modes in the signal, and the calculation here assumes that all of the wave energy, without distinction as to mode, is delivered to particles on the time scale of ion acoustic damping. This leads to an unavoidable overestimate. As a partial indication of the presence of other modes, the correlation between the magnitude of the magnetic field and the density has been calculated for each 389 sec set. A negative correlation is generally taken as representing ion acoustic waves. For the low pressure set, of the correlations 66 were negative and 62 were positive. For the high pressure set the numbers were, oddly, the same. This is taken as indication that mixing of other modes is considerable but that appreciable fraction of the wave energy is due to ion acoustic waves. Of the three causes of the overestimate, this mixing is probably the least important.

Second, it is interesting to investigate the very large peaks seen, far above the observed heating, especially in the low pressure set. In figure 5 are shown some plasma parameters for two large peaks. On the left are parameters for the peak at 7.8 hours in Figure 4. For this large peak, the density-magnetic field correlation is negative, -.63, as can be seen. indicating a major component of ion acoustic waves. On the right are the parameters of the largest peak at 3.45 hours. The density-magnetic field correlation is positive, +.76, indicating a large component of shear Alfven waves. These are both boundaries between two different plasmas. There is no other easy distinguishing characteristic of the two, and it seem that the peaks are simply due to very large wave energy that is a consequence of the large pressure change. This is just the situation that it was attempted to avoid by choosing short sections to be analyzed, but as is seen, it was sometimes not successful. The program automatically assigns a heating of f/4 to these, but of course, the energy is not due to waves that are the interest of this work.

Third, in the calculation of heating the expression for damping has been used which is appropriate for wave propagation parallel to the magnetic field. As pointed out by Narita and Marsch (2015), the damping of ion acoustic waves becomes much less for oblique propagation. In Figure 6 is shown the ratio of imaginary part of the frequency to the real part as a function of theta, the angle between the wave vector and the magnetic field, from a Vlasov calculation. The plasma parameters have been taken from a period in the middle of the low pressure set, and for a frequency midway between the extremes of a 389 sec period. It has been known for a long time that the cascade favors strongly oblique daughters of the cascade processes (Sridhar and Goldreich, (1994), Goldreich and Sridhar, (1995)). The lowest damping in Figure 6 is $-\omega i/\omega r = .011 = 1/91$. If $\omega/91$ is used to the damping in Figure 5 instead of $\omega/4$, the heating, except for the highest peaks, falls below the observations. Figure 7 shows a recalculation, using this slower damping, of the high pressure study, the worst discrepancy. It will be seen that the allowed range of oblique damping would allow an obliquity giving full agreement with the observations.

There are then three causes of the overestimates shown in Figure 4, mixing of other modes, mixing of different plasmas, and reduced damping of oblique ion acoustic waves. In accounting for these, it seems that the damping of ion acoustic-kinetic slow mode waves can provide most or nearly all of the observed heating.

### 3.1.1 Wave Energy from Electric Field Spectra

The second term in the Rayleigh formula, Eq. 3, might also be used to evaluate the heating. The particles velocities are due to the electric fields of the waves. Accordingly:

$$dv/dt = eE/m \qquad \delta v = eE(\omega)/(m\,\omega)$$

These two parts of the second term, the kinetic energy of the different particles, is clearly dominated by the electron part, so drop Ions in what follows.

$$dv/dt = eE/m_e \qquad \delta v = eE(\omega)/(m_e \omega) \tag{6}$$

giving for the energy per unit volume and frequency range:

$$dW_j(\omega,k)/dVd\omega = (n/2m_e)(eE(\omega)/\omega)^2 = 7.\,10^{-2} <E(\omega)/\omega>^2 \; J/m^3 \; \delta\omega \tag{7}$$

In this, the particle density for each species has been taken as $5.\,10^6$ in correspondence with other evaluations used in this work.

In the ion acoustic-slow mode, the ion and electron densities are very close to equal and the electric fields are small. Consequently, the energy of the particles is considerably larger than that of the fields, and the coefficient of $<E^2>$ in the energy is large. The coefficient of the electric field energy is then:

$$2(n/2m_j)(e/\omega)^2/\varepsilon_0 \qquad = 1.6\,10^{10}/\omega^2 \tag{8}$$

This is, as stated above, very large and increases toward lower frequency

Measurements of the electric field spectrum are usually reported as:

$$<E^2(f_{obs})> = A_E \, f^{-\alpha} \; (V/m)^2/Hz$$

The wave energy spectral density $dW(\omega,k)/dVd\omega$ is then

/

$$dW_j(\omega,k)/dV\text{-}df = (n\,e^2/2m_e)((1/2\pi)\,V_{sw}/V_s)^{-\alpha-2}(A_E\,f^{-\alpha-2}) = .6\,(A_E\,f^{-\alpha-2})$$

In this, the $V_{sw}/V_s$ etc factor has been evaluated for $a = 5/3$ and the subscript on f dropped. The total energy is then obtained by integrating over the frequency spectrum over a range from $f_{low}$ to $f_{high}$. If $f_{high}$ is reasonably far above $f_{low}$, it may be ignored. However, what is desired here is the heating rate, i.e. the rate at which this energy is transferred to the plasma. This rate is given by the rate of decay of ion acoustic waves, approximately 1/3 to 1/2 of the real part of the frequency, $\omega = f/1.8$, so that the heating, H, in joules per $m^3$ per sec, again integrating over the frequency spectrum, is:

$$H = .22\,(A_E\,f_{low}^{-\alpha-1})$$

Kellogg et al (2006) found $A_E\,f^{-\alpha} = 10^{-10}\,f^{-5/3}$ $(V/m)^2/Hz$, Bale et al (2006) found $8.\,10^{-10}\,f^{-5/3}$.

For the Kellogg et al spectrum: $H = 6.\,10^{-11}\,f_{low}^{-\alpha-1}$ J/m3-sec

This result depends, as can be seen, critically on the lower limit of the ion acoustic wave spectrum. Published measurements of the electric field spectrum do not show any break that might be interpreted as a change of mode. For an example, Chen et al (2011), show a spectrum with a break at about $2\,10^{-5}$ Hz which corresponds roughly to the break in the magnetic field spectrum corresponding to the lower limit of the inertial region. A more direct measurement that might be interpreted as the lower limit of ion acoustic waves could come from Vellante and Lazarus (1987) who investigated the correlation between B and density down to $1.6\,10^{-4}$ Hz (period 2 hours). A lower estimate due to Goldman and Siscoe (1972) is quoted by Tu and Marsch(1995) p 120, as between and $4\,10^{-5}$ (7 hours). However, even if the rather high lower limit of 2.6 10-3 Hz corresponding to the lower limit of the 389 sec samples used for the pressure heating, the result is enormously greater than the observations. It seems that this large electric field must come from a part of the electric field spectrum due to shear Alfven waves, which have only a very slow decay, especially at low frequencies, and so do not contribute much to the heating. It also suggests that the calculations above from pressure measurement may sometimes be too large because there may be some Alfven waves in the spectrum However, this estimate from the second Raayleigh term is much larger than that from the pressure term and must be dominated by other waves.

In the Rayleigh expression, Eq. (3), the two terms are supposed to alternate, with equal maxima. This is manifestly not the case here. In air, there is only one wave mode, while in plasma there are several. It seems that the two terms can emphasize different modes and that the second term, the velocity energy, must emphasize the modes other than the ion-acoustic mode. The other modes must also play some role in the first term.

**From Turbulence**

It has been generally thought that turbulent magnetic fields in the dissipation region account for the heating, and a long set of papers, (e.g.Usmanov, et al 2011) and references therein) has provided evidence for this conclusion.. The assignment of the density fluctuation fraction of the turbulence to the ion-acoustic mode implies that at least part of the heating is due to absorption of the energy of these waves, with a further implication that the energization does not all take place at the high frequency end of the inertial range. The present work does not alter this. The ion acoustic waves are very evanescent and some process must replace them rapidly. They are then just a step in the heating process. Such a process has been suggested by Derby (1978), Goldstein(1978) and Bowen et al (2018), but it has not been shown that their process generates enough waves. The nonlinear process of Derby and Goldstein are a nonlinear decay of waves propagating parallel to the magnetic field. Matteini et al (2010) in simulations have shown that oblique waves also lead to heating, but simulations do not allow the same detailed examination of the process, including growth rates and the nature of daughter products, that algebraic calculations do.

**Conclusions**

It seems that, within the uncertainties of these calculations, the heating of the solar wind by absorption of ion acoustic-kinetic slow mode waves is a significant part and perhaps nearly all of the observed heating of the solar wind in the regions reached before the Parker Solar Probe mission. This verifies the early suggestions of Marsch and Tu (1994,1995), and Howes et al; (2012) that the ubiquitous density fluctuations are in the ion acoustic-kinetic slow mode and the decay of these provides heating of the solar wind. This does not contradict the longstanding belief that the heating is due to magnetic field turbulence. (For a review of the enormous literature on this subject see Usmanov et al (2011)). The ion acoustic waves are a step in the conversion of turbulence to heat, and the generation of these short duration waves from the turbulence remains to be understood. A suggestion, but at higher frequency and in the fast wind, was made by Jiling (1998). Some progress has been made, algebraically by Derby(1978) and Goldstein(1978), observationally by Bowen et al (2108), and in simulations by Matteini et al (2010) , but understanding is not yet complete.

**Acknowledgements**

This work was supported by the National Aeronautics and Space Administration under grant NNX14AK73G.

The author thanks CDAWeb and S.D.Bale and R.P.Lin for the pressure and plasma density data. Thanks to CDAWeb and R.Lepping for the magnetic field data. Thanks also to Greg Howes, (University of Iowa)*,* Kris Klein (now at University of Arizona) and Marco Velli (University of California at Los Angeles) for helpful discussions. Thanks also to OMNIweb plus, Natalia Papitashvili and Marcia Neugebauer for the solar wind speed data from the Mariner II mission.

The data used in this work are available to the public either from the NSSDC or from published data.

Figure Captions

Figure 1.  Histograms of the Ratios of Magnetic Pressure to Particle Pressure for Two Periods

Figure 2.  The Spectrum of Wave Energy for the High Ion Pressure period.
The red line represents the spectrum as $f^{(-.80)}$

Figure 3.  Particle temperature vs density for the high pressure and the low pressure periods.

Figure 4. The heating rates, in Joules/m3-sec for the two periods studied, together with the heating rates found by Coleman(1968) near 1 AU and Gazis and Lazarus (1982) beyond 1 AU. as red lines.

Figure 5  Pressure, density, temperature and magnetic field for two heating peaks.  Also shown is a line for the average pressure, important in calculating the wave energy.

Figure 6.  Ratio of the imaginary part of the frequency to the real part as a function of angle theta between the wave vector and the magnetic field.  Plasma parameters are for a period during the low pressure set.

Figure 7.  Calculated and observed heating for a section of the high pressure set, using the lowest damping from Figure 6

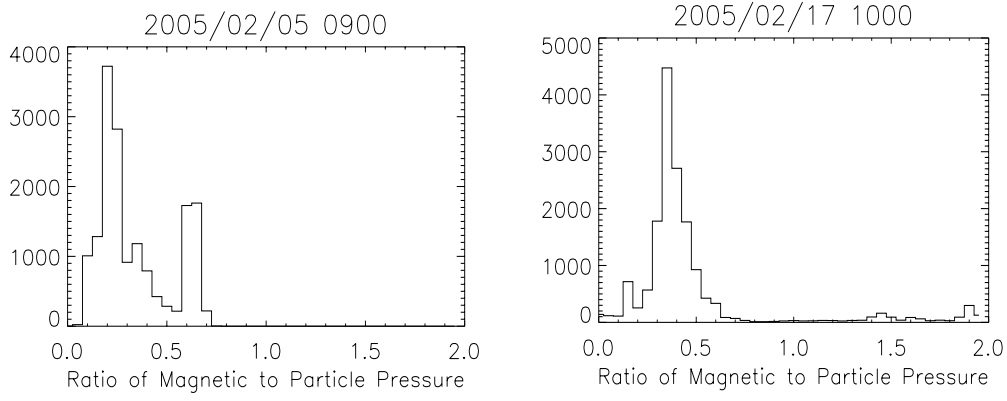

Figure 1. Histograms of the Ratios of Magnetic Pressure to Particle Pressure for Two Periods

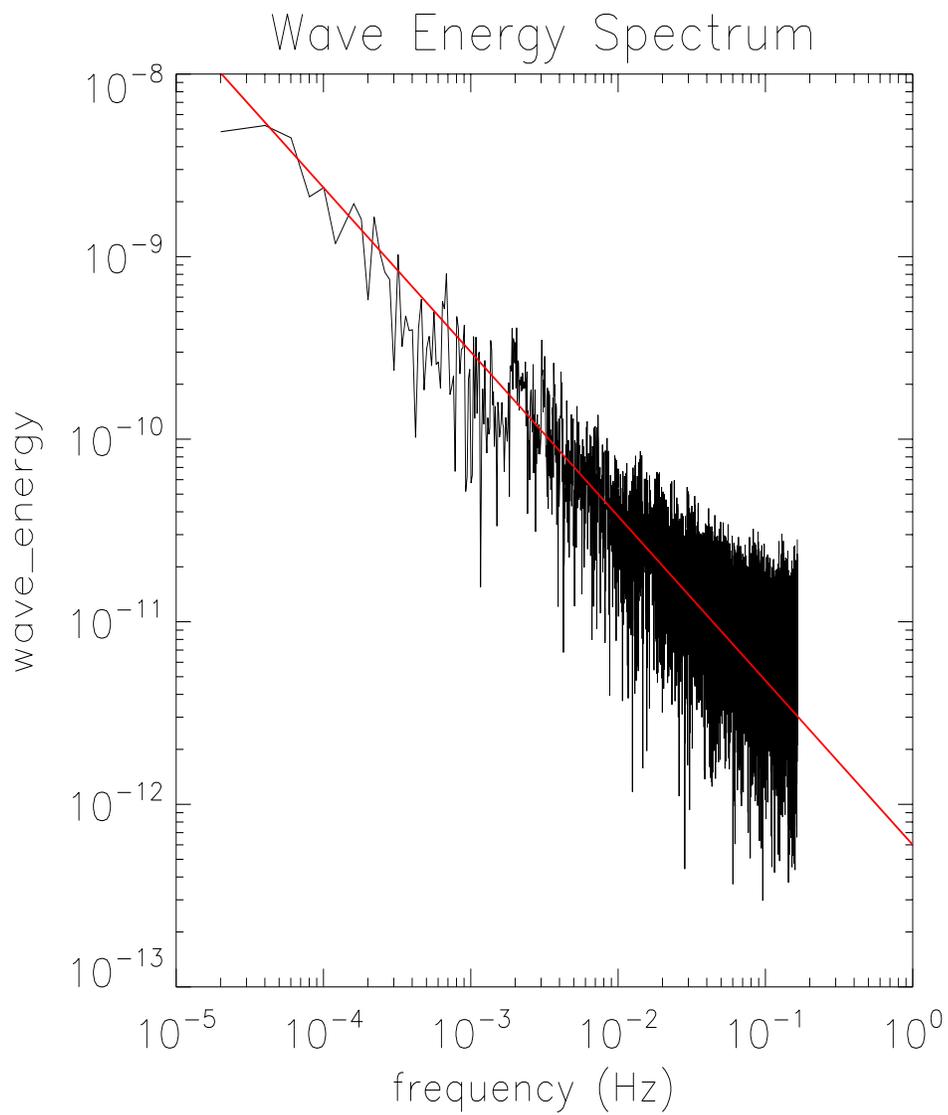

Figure 2. The Spectrum of Wave Energy for the High Ion Pressure period. The red line represents the spectrum as $f^{(-.80)}$

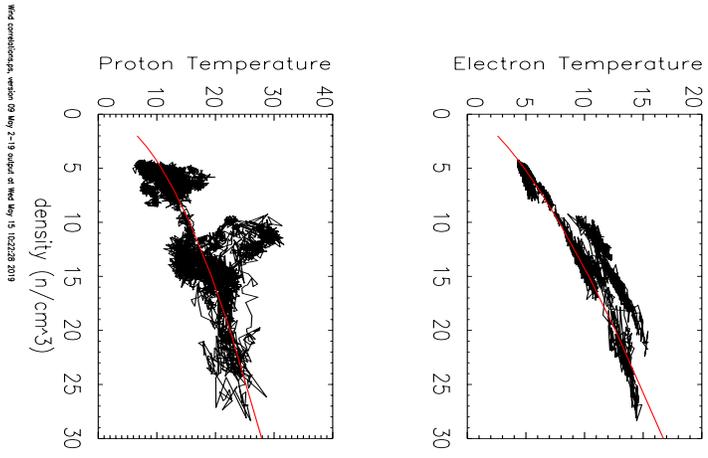

Figure 3. Particle temperature vs density for the high pressure and the low pressure periods.

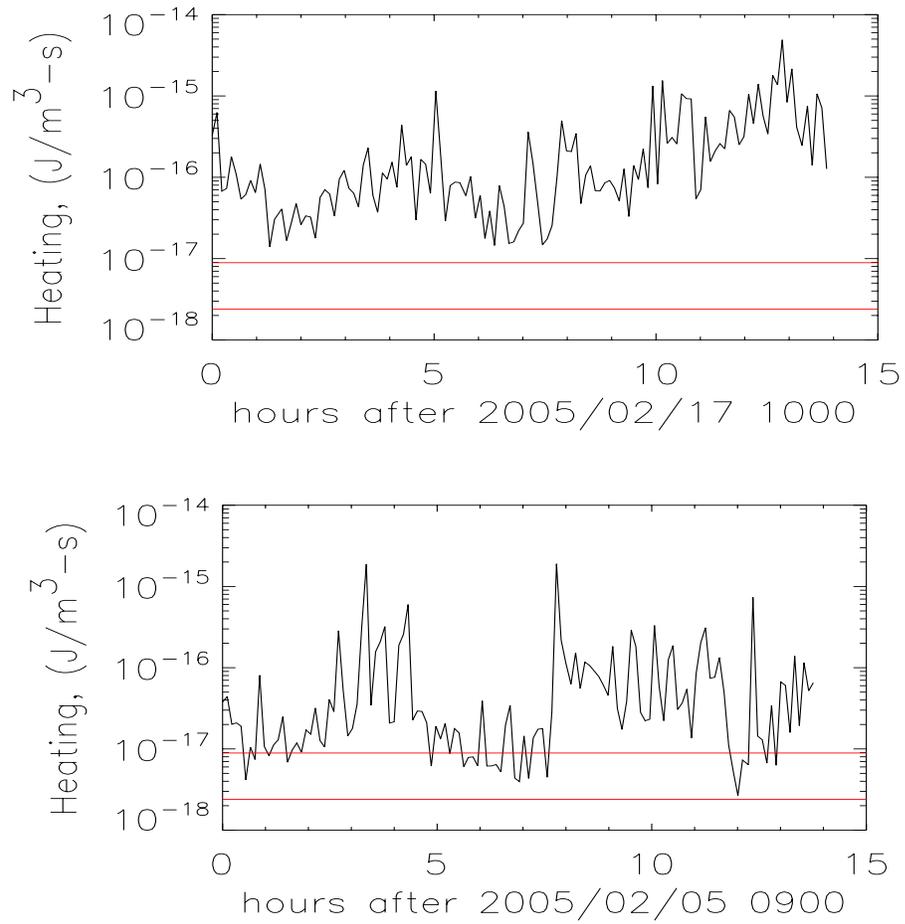

Figure 4. The heating rates, in Joules/m3-sec for the two periods studied, together with the heating rates found by Coleman(1968) near 1 AU and Gazis and Lazarus (1982) beyond 1 AU. as red lines.

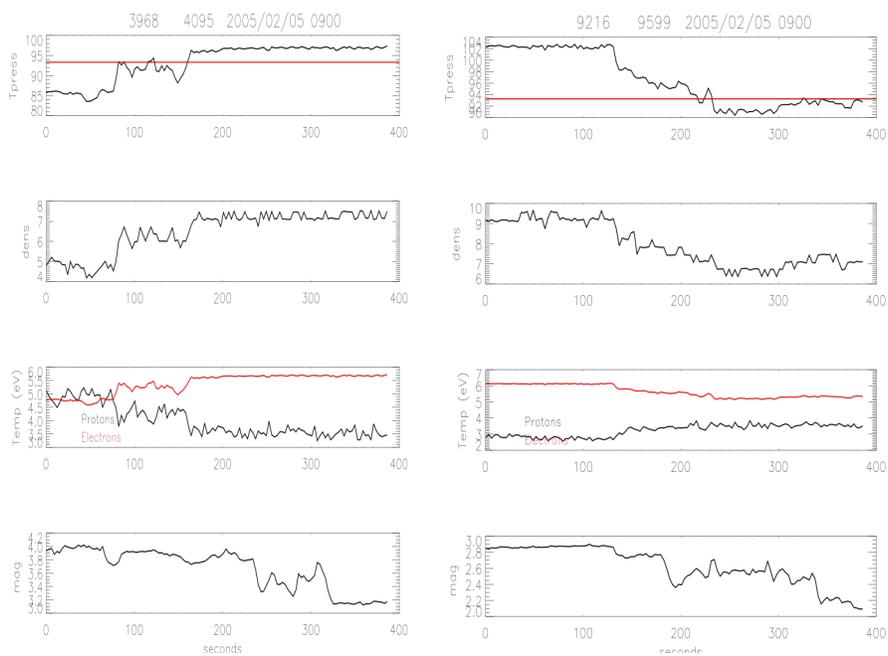

Figure 5  Pressure, density, temperature and magnetic field for two heating peaks.  Also shown is a line for the average pressure, important in calculating the wave energy.

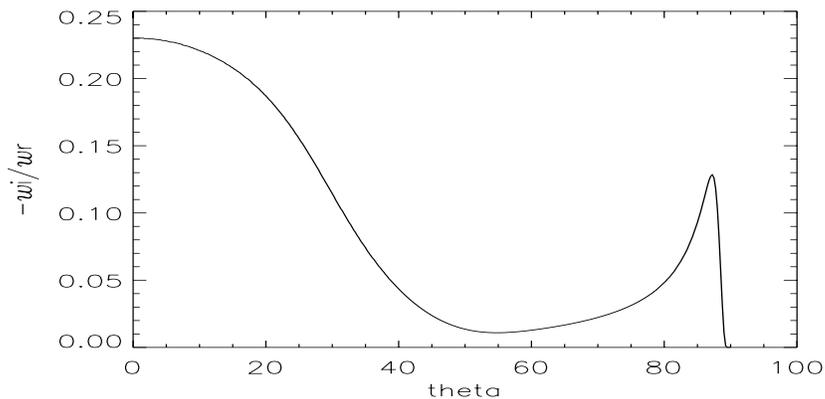

Figure 6.  Ratio of the imaginary part of the frequency to the real part as a function of angle theta between the wave vector and the magnetic field.  Plasma parameters are for a period during the low pressure set.

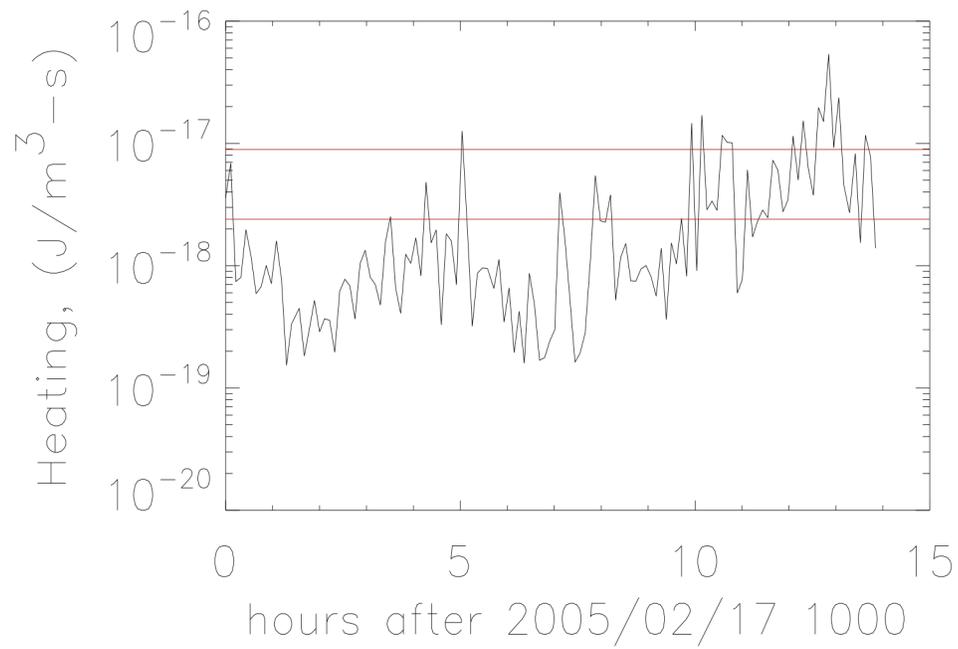

Figure 7. Calculated and observed heating for a section of the high pressure set, using the lowest damping from Figure 6